\documentclass[11pt,reqno]{amsart}
\usepackage{amsfonts}
\usepackage{mathrsfs}
\usepackage{amsmath}
\usepackage{amssymb,amsfonts}

\newtheorem{thm}{Theorem}[section]

\newtheorem{prop}{Proposition}[section]
\numberwithin{equation}{section}
\newtheorem{rmk}{Remark}[section]

\def\pf{{\textit {Proof:} }}

\newcommand{\mysection}[1]{\section{#1}\setcounter{equation}{0}}

\newfont{\bb}{msbm10 at 11pt}

\def\d{\mathrm{d}}
\def\I{\mathrm{i}}

\def\d{\mathrm{d}}

\def\ban#1\ean{{\allowdisplaybreaks \begin{align}#1\end{align}}}
\def\baN#1\eaN{{\allowdisplaybreaks \begin{align*}#1\end{align*}}}

\newcommand{\bal}{\begin{aligned}}      \newcommand{\eal}{\end{aligned}}
\newcommand{\ba}{\begin{array}}      \newcommand{\ea}{\end{array}}
\newcommand{\bc}{\begin{center}}     \newcommand{\ec}{\end{center}}
\newcommand{\be}{\begin{enumerate}}  \newcommand{\ee}{\end{enumerate}}
\newcommand{\beq}{\begin{eqnarray}}  \newcommand{\eeq}{\end{eqnarray}}
\newcommand{\beQ}{\begin{eqnarray*}} \newcommand{\eeQ}{\end{eqnarray*}}
\newcommand{\bi}{\begin{itemize}}    \newcommand{\ei}{\end{itemize}}
\newcommand{\bt}{\begin{tabular}}    \newcommand{\et}{\end{tabular}}
\newcommand{\bdm}{\begin{displaymath}} \newcommand{\edm}{\end{displaymath}}

\newcommand{\lv}{\left\vert}
\newcommand{\rv}{\right\vert}

\newcommand{\wt}{\widetilde}
\newcommand{\lb}{\left(}
\newcommand{\rb}{\right)}
\newcommand{\lba}{\left(\begin{array}}      \newcommand{\ear}{\end{array}\right)}

\def\qed{\hfill{Q.E.D.}\smallskip}

\newcommand{\ls}{\setlength{\baselineskip}{12pt}
	\setlength{\parskip}{3mm}} 

\begin{document}
	
	\title[]{The Dirac and  Rarita-Schwinger equations on scalar flat metrics of Taub-NUT type}
	
	\author[X Xue]{Xiaoman Xue$^{\dag}$}
	\address[]{$^{\dag}$ School of Physical Science and Technology, Guangxi University,  Guangxi 530004, China}
	\email{xuexiaoman@st.gxu.edu.cn}
	\author[C Liu]{Chuxiao Liu$^{\flat}$}
	\address[]{$^{\flat}$  School of Mathematics and Information Science, Guangxi University,  Guangxi 530004, China;}
	\address[]{$^{\flat}$ Center for Mathematical Research, Guangxi University,  Guangxi 530004, China;}
	\address[]{$^{\flat}$ Guangxi Base, Tianyuan Mathematical Center in Southwest China, Guangxi 530004, China}
	\email{cxliu@gxu.edu.cn}
	
	\date{}
	
	\begin{abstract}
	     We construct a scalar flat metric of Taub-NUT type whose total mass can be negative.
	     The standard Taub-NUT metric and its negative NUT charge counterpart serve as particular examples, for which the complex 2-dimensional space of parallel spinors gives rise to $L^2$ harmonic spinors and Rarita-Schwinger fields.
	     For the scalar flat Taub-NUT type metric, we study the Dirac and Rarita-Schwinger equations by separating them into angular and radial equations, and obtain explicit solutions in certain special cases. 
	\end{abstract}
	
	\maketitle \pagenumbering{arabic}
	
	\mysection{Introduction}\ls
	
	Let  $\theta$, $\phi$, $\psi$ be the Euler angles on the $3$-sphere $S^3$, with the Cartan-Maurer one-forms given by 
	\begin{align*}
		\sigma_{1} &=\sin\psi \d\theta- \sin\theta \cos\psi \d\phi,\\
		\sigma_{2} &=-\cos\psi \d\theta- \sin\theta \sin\psi \d\phi,\\
		\sigma_{3} &=\d\psi+ \cos\theta \d\phi.
	\end{align*}
   A metric of Taub-NUT type is defined by 	
	\begin{align}\label{TNT}
		g=f^2 (r)\d r^2+ \lb r^2- N^2 \rb  \lb \sigma_{1}^2+
		\sigma_{2}^2 \rb + 4N^2 f^{-2} (r) \sigma_{3}^2.
	\end{align}
	It is referred to as the Taub-NUT metric when
	\begin{align}\label{TN}
		f(r)= \sqrt{ \frac{r+N}{r-N} }
	\end{align}
	with the ranges
	\begin{align*}
		r \geq N > 0, \quad  0 \leq  \theta  < \pi,\quad 0 \leq  \phi  < 2 \pi, \quad  0 \leq \psi < 4\pi. 
	\end{align*}
	This metric is a complete, Ricci flat  Riemannian metric on $\mathbb{R}^4$ that appears as one of the gravitational instantons and plays an important role in the Euclidean approach to quantum gravity \cite{EGH, GH, H}.  
	It is also referred to as the Taub-NUT metric with  negative NUT charge  when
	\begin{align}\label{TN-N}
		f(r) = \sqrt{\frac{r - N}{r + N}}.
	\end{align}
	This metric is Ricci flat, has a curvature singularity at $r = N$, and coincides with the asymptotic form of the Atiyah-Hitchin metric \cite{AH1}. 
	
	For mathematical and physical interest, the Dirac equation has been extensively studied on gravitational instantons  (see, e.g., \cite{CZ, CXZ, F2, F, P, SU} and references therein). 
    Harmonic spinors are defined to be the zero modes of the Dirac equation, and their existence is related to the topological properties of the background geometry. 
    In particular, in \cite{CZ},  explicit  harmonic spinors were obtained via separation of variables on  scalar flat metrics of Eguchi-Hanson type, such metrics were constructed  in \cite{Z}  by solving an ordinary differential equation. 
	In this paper, we construct a scalar flat metric \eqref{TNT} of Taub-NUT type  where
	\begin{align}\label{TNTf}
		f(r)=\sqrt{\frac{r^2-N^2}{r^2 +C_1 r +C_2}}
	\end{align}
	with the constants $C_1$ and $ C_2$ satisfying $C_1\geq -2N$ and 
	\begin{align*}
		-N^2 -N C_1	\leq	C_2 \leq \frac{(C_1)^2}{4}.
	\end{align*}
	The metric is geodesically complete for $r \geq N$ when  $C_2 = -N^2 -N C_1$,  whereas it develops a curvature singularity at $r = N$ when $C_2 > -N^2 -N C_1$.
	As the total energy can be negative for the scalar flat Taub-NUT type metric, this motivates us to study the Dirac equation to understand why Witten's spinorial approach does not work for the positive energy theorem.
	
	The Rarita-Schwinger equation for $\frac{3}{2}$-spinors is a generalization of the Dirac equation for $\frac{1}{2}$-spinors.
	It was first introduced by Rarita and Schwinger \cite{RS} and is of great importance in supergravity and superstring theories. 
	This equation has been studied using separation of variables on Kerr spacetime \cite{G1, TS} and, more generally, on type-D vacuum backgrounds \cite{K2}.
	Rarita-Schwinger fields are zero modes of this equation and are divergence-free (see, e.g., \cite{AE, BH, HS, OT, W} and references therein).
	In \cite{AE}, A\c{c}\i k and Ertem  constructed Rarita-Schwinger fields on Ricci flat metrics from spin-$1$ Maxwell fields using twistor spinors.
	However, their construction is not applicable to the metric \eqref{TNT} with $f(r)$ given by \eqref{TNTf}, since the metric has no nontrivial twistor spinors. 
	This motivates us to study the Rarita-Schwinger equation on \eqref{TNT} via separation of variables.

	In this paper, we show that for the metric \eqref{TNT} with $f(r)$ given by \eqref{TN} or \eqref{TN-N}, the spaces of twistor and parallel spinors coincide and are complex 2-dimensional. 
    Using these parallel spinors together with harmonic functions or Maxwell fields, we obtain $L^2$ harmonic spinors for \eqref{TN-N}, and $L^2$ Rarita-Schwinger fields for both \eqref{TN} and \eqref{TN-N}.
    We then separate  the  Dirac equation on the metric \eqref{TNT} into angular and radial equations. 
	For $\lambda=0$ and $f(r)$ given by \eqref{TNTf} with $C_2 > -N^2 -N C_1$, explicit $L^2$ solutions to the radial equations on $r>N$ are obtained by direct integration.
	For $\lambda \neq 0$ and $f(r)$ given by \eqref{TN-N},  explicit solutions  on $r>N$ are expressed via Kummer functions. 
	We also separate the Rarita-Schwinger equation on \eqref{TNT}.  Similarly,  the radial equations  for $\lambda=0$ and  $f(r)$  given by \eqref{TNTf} with $C_2 > -N^2 -N C_1$ are solved,  yielding solutions  on $r\geq N$ that are $L^2$ integrable.
	For $\lambda \neq 0$ and $f(r)$ given by \eqref{TN},  explicit solutions on $r\geq N $ are obtained in terms of Kummer functions.

	This paper is organized as follows. 
	In Section 2, we construct a scalar flat Taub-NUT type  metric \eqref{TNT} and show that its total mass can be negative. We also investigate the almost-complex structures on \eqref{TNT}.
	In Section 3, we prove that the spaces of twistor and parallel spinors coincide and are complex 2-dimensional for $f(r)$ satisfying \eqref{TN} or \eqref{TN-N}. Furthermore, we provide the Dirac and Rarita-Schwinger equations on \eqref{TNT}.
	In Section 4, we obtain $L^2$ harmonic spinors and  Rarita-Schwinger fields  using the parallel spinors.
	In Section 5, we separate the Dirac equation and solve the angular and radial  equations in certain cases.
	In Section 6, we separate the Rarita-Schwinger equation and solve the angular and radial equations in certain cases.
			
\mysection{Scalar flat metrics of Taub-NUT type}\ls
  In this section, we focus on constructing a scalar flat metric of the form \eqref{TNT}, where $f$ is uniquely given by \eqref{TNTf}, and proving that its total mass can be negative.
   
  Let coframe of \eqref{TNT} 
	\begin{align*}
		e^{1}= f \d r,\quad 
		e^{2}= \sqrt{r^2 -N^2} \sigma_{1},\quad
		e^{3}= \sqrt{r^2 -N^2} \sigma_{2},\quad
		e^{4}= 2N f^{-1} \sigma_{3},
	\end{align*}
	and dual frame  
	\begin{align*}
		&e_{1}=\frac{1}{f}\partial_{r},\quad
		e_{2}=\frac{1}{\sqrt{r^2 -N^2}} \lb \sin\psi \partial_{\theta} - \frac{\cos\psi}{\sin\theta} \partial_{\phi} + \frac{\cos\theta\cos\psi}{\sin\theta} \partial_{\psi}\rb,\\
		&e_{3}=\frac{1}{\sqrt{r^2 -N^2}} \lb -\cos\psi \partial_{\theta} - \frac{\sin\psi}{\sin\theta} \partial_{\phi} + \frac{\cos\theta\sin\psi}{\sin\theta} \partial_{\psi}\rb, \quad
		e_{4}=\frac{f }{2N}\partial_{\psi}.\\
	\end{align*}
The connection 1-form ${\omega^{\alpha}} _{\beta}$ and the curvature 2-form ${R^\alpha}_{\beta}$ are defined by
\begin{align*}
	\d e^{\alpha}+{\omega^{\alpha}}_{\beta}\wedge e^{\beta}=0, \quad {R^\alpha}_{\beta}={\d\omega^\alpha}_{\beta}+{\omega^\alpha}_{\gamma}\wedge{\omega^\gamma}_{\beta}.
\end{align*}
The nonzero components are given as follows
	\begin{align}	\label{w}
		\begin{split}
			{\omega^{2}}_{1}&=\frac{r}{\lb r^2-N^2 \rb f} e^2,\quad
			{\omega^{3}}_{1}=\frac{r}{\lb r^2-N^2 \rb f} e^3,\\
			{\omega^{4}}_{1}&=-\frac{f'}{f^2} e^4,\qquad\qquad\,
			{\omega^{3}}_{2}=\lb\frac{N}{\lb r^2-N^2 \rb f}-\frac{f}{2N}\rb e^4,\\
			{\omega^{4}}_{2}&=\frac{N}{\lb r^2-N^2 \rb f} e^3,\quad
			{\omega^{4}}_{3}=-\frac{N}{\lb r^2-N^2 \rb f} e^2,
		\end{split}
	\end{align}
and
	\begin{align*}
		{R^{2}}_{1}&=A(r) e^2 \wedge e^1 - B(r) e^4 \wedge e^3,\\
		{R^{3}}_{1}&=A(r) e^3 \wedge e^1 + B(r) e^4 \wedge e^2,\\
		{R^{4}}_{1}&=\frac{f f''- 3 \lb f'\rb^2}{f^4} e^4 \wedge e^1 + 2B(r) e^3 \wedge e^2,\\
		{R^{3}}_{2}&=2B(r) e^4 \wedge e^1 + \frac{\lb r^2-N^2 \rb f^2 - 3N^2 -r^2}{\lb r^2-N^2 \rb^2 f^2} e^3 \wedge e^2,\\
		{R^{4}}_{2}&=B(r) e^3 \wedge e^1 + A(r) e^4 \wedge e^2,\\
		{R^{4}}_{3}&=-B(r) e^2 \wedge e^1 + A(r) e^4 \wedge e^3,
	\end{align*}
	where 
	\begin{align*}
		A(r)=\frac{N^2 f + r \lb r^2-N^2 \rb f'}{\lb r^2-N^2 \rb^2 f^3},\quad B(r)=\frac{N r f+N \lb r^2-N^2 \rb f'}{\lb r^2-N^2 \rb^2 f^3}.
	\end{align*}
	Therefore, the scalar curvature of the metric \eqref{TNT} is
	\begin{align*}
		R=-\frac{2\lb f^2 -f^4 - 4r f f' -\lb r^2-N^2 \rb f f''+3 \lb r^2-N^2 \rb \lb f'\rb^2 \rb}{\lb r^2-N^2 \rb f^4}.
	\end{align*}
	Let $f=\sqrt{ \frac{r^2-N^2}{h} }$. The resulting  equation for scalar flat metrics is
	\begin{align*}
		h'' - 2=0.
	\end{align*}
	Its solutions  are
	\begin{align*}
		h=r^2 + C_{1}r + C_{2}
	\end{align*}
	for some real constants  $C_1$, $C_2$. Thus, the metric \eqref{TNT} is scalar flat for $f$ given by  \eqref{TNTf}. 
 		
 	The nontrivial Ricci curvature components are
 	\begin{align}\label{Ricci}
 		R_{11}=R_{44}=\frac{N^2 - C_{2}}{\lb r^2-N^2 \rb^2},\quad R_{22}=R_{33}=\frac{C_{2} - N^2}{\lb r^2-N^2 \rb^2}.
 	\end{align}
 	When $C_1 = \pm 2N$ and $C_2=N^2$, the metric reduces to the Taub-NUT metric and its negative NUT charge  counterpart, both of which were noted by Atiyah and Hitchin to be hyperk\"ahler \cite[Chap. 9]{AH}. Using the method of  \cite{K3}, we study the metric \eqref{TNT}  by rewriting its coframe  as follows:
 	\begin{align*}
 			\wt{e}^1&= f \sin\theta \cos\phi \d r +\sqrt{r^2 -N^2} \lb \cos\theta \cos\phi \d \theta -\sin\theta \sin\phi \d \phi \rb, \\
 			\wt{e}^2&= f \sin\theta \sin\phi \d r +\sqrt{r^2 -N^2} \lb \cos\theta \sin\phi \d \theta +\sin\theta \cos\phi \d \phi \rb, \\
 			\wt{e}^3&= f \cos\theta \d r -\sqrt{r^2-N^2} \sin\theta  \d \theta, \\
 			\wt{e}^4&= \frac{2 \delta N }{f} \lb \d \psi + \cos\theta \d \phi \rb,
 		\end{align*}
 	where $\delta=\pm 1$. With respect to this coframe, we define three almost-complex structures
 	\begin{align}\label{TNTJ}
 		\begin{split}
 			J_1(\wt{e}^1,\; \wt{e}^3)=(\wt{e}^2, \; \wt{e}^4),\\
 			J_2(\wt{e}^1,\; \wt{e}^4)=(\wt{e}^3, \; \wt{e}^2), \\
 			J_3(\wt{e}^1,\; \wt{e}^2)=(\wt{e}^4, \; \wt{e}^3).	
 		\end{split}
 	\end{align}
 	\begin{prop}
 		The three almost-complex structures $J_1, J_2, J_3$ given by \eqref{TNTJ} are not integrable unless $f$ satisfies \eqref{TN} or \eqref{TN-N}.
 	\end{prop}
 	\pf For the almost-complex structure $J_1$,  the $(1,0)$-forms are given by
 	\begin{align*}
 		\omega_1=\wt{e}^1 +\I \wt{e}^2,\quad
 		\omega_2=\wt{e}^3 + \I \wt{e}^4.
 	\end{align*}
 	A direct computation yields
 	\begin{align*}
 		\d\omega_1 = & \lb \frac{r}{r^2-N^2}-\frac{1}{\sqrt{r^2 +C_1 r +C_2}}\rb \d r \wedge \omega_1, \\
 		\d\omega_2 = & - \frac{ 2 \delta N   e^{-\I\phi}}{f \sqrt{r^2 -N^2} }\d \theta \wedge \omega_1 -\frac{f'}{f} \d r \wedge \omega_2\\ 
 		&+\frac{Q(r)}{2\sqrt{r^2-N^2}(r^2 +C_1 r +C_2)}\sin\theta \d r \wedge\d \theta,
 	\end{align*}
 	where
 	\begin{align*}
 		Q(r)= & 2(r^2-N^2)\sqrt{r^2 +C_1 r +C_2}\\ & -(r+ \delta N)(2 r^2 +3C_1 r +2\delta N r +\delta N C_1 +4C_2).
 	\end{align*}
 The integrability of $J_1$​ requires that $\omega_1$ and $\omega_2$ be closed, i.e., $\d\omega_1=0$ and $\d\omega_2=0$, which implies  
 	\begin{align*}
 		C_1=-2\delta N, \quad C_2=N^2.
 	\end{align*}
 	A similar analysis shows that $J_2$ and $J_3$ are also integrable when the above conditions hold, and this is the Taub-NUT metric and  its negative NUT charge  counterpart. 
 	\qed

    The total mass of the scalar flat metric \eqref{TNT} is computed below. Let
	\begin{align} \label{g0}
		\breve{g}= 	\d r^2+(r^2-N^2)\lb \sigma_{1}^2 +\sigma_{2}^2 \rb +4N^2 \sigma_{3}^2
	\end{align}
	be the flat metric,
	\begin{align*}
		\breve{e}^{1}=\d r, \quad \breve{e}^{2}=\sqrt{r^2-N^2} \sigma_{1}, \quad  \breve{e}^{3}=\sqrt{r^2-N^2}\sigma_{2}, \quad \breve{e}^{4}=2N\sigma_{3}
	\end{align*}
	be its coframe and $\{\breve{e}_{a}\}$ be its frame. The total mass of the scalar flat  Taub-NUT type metric is 	
	\begin{align*}
		E= \frac{1}{4 \text{vol}(S^3)} \lim_{r\to \infty} \int_{\partial M_{r}} \lb \breve{\nabla}^{j} g_{ij} -\breve{\nabla}_{i} \text{tr}_{
			\breve{g} }(g)\rb \star \breve{e}^{i},
	\end{align*}
	where $g_{ij}=g(\breve{e}_{i}, \breve{e}_{j})$, $\breve{\nabla}$ and $\star$ are the Levi-Civita connection and  the Hodge star operator of the metric \eqref{g0}. 
	The connection 1-forms of \eqref{g0} are
	\begin{align*}
		{\breve{\omega}^2}_{\;\;1}= \frac{r}{r^2-N^2}& \breve{e}^2,\quad
		{\breve{\omega}^3}_{\;\;1}= \frac{r}{r^2-N^2} \breve{e}^3,\quad
		{\breve{\omega}^2}_{\;\;3}= \lb\frac{1}{2N} -\frac{N}{r^2-N^2} \rb\breve{e}^4,\\
		&{\breve{\omega}^2}_{\;\;4}	= -\frac{N}{r^2-N^2} \breve{e}^3,\quad
		{\breve{\omega}^3}_{\;\;4}= \frac{N}{r^2-N^2} \breve{e}^2.
	\end{align*}
  Therefore, 
	\begin{align*}
		\breve{\nabla}^{j} g_{1j}-\breve{\nabla}_{1} \text{tr}_{
			\breve{g} }(g) &=\breve{e}_{j}(g_{1j}) -g_{1l}{ \breve{\omega}^{l} }_{\; j}(\breve{e}_{j}) -g_{jl}{ \breve{\omega}^{l} }_{\;1}(\breve{e}_{j})- 2f f' + \frac{2f'}{f^3}\\
		& =\frac{2r \lb f^2 -1 \rb}{r^2-N^2} + \frac{2f'}{f^3}.
	\end{align*}
	Denote the domain
	\begin{align}\label{D}
		\mathcal{D}=\left\{0 \leq \theta < \pi, \; 0 \leq \phi < 2\pi, \; 0 \leq \psi < 4 \pi \right\},
	\end{align}
	it follows that
	\begin{align*}
		V_{0}=\int_{\mathcal{D}} \sigma_{1}\sigma_{2}\sigma_{3}=16\pi^2.
	\end{align*}
	We obtain
	\begin{align*}
		E=\frac{-2N C_{1}}{ 4 \text{vol}(S^3)} V_{0}=-4N C_{1},
	\end{align*}
    which is negative for $C_1>0$.

	\mysection{The Dirac and Rarita-Schwinger equations} 
In this section, we study the twistor spinors and present the Dirac and Rarita-Schwinger equations on the metric \eqref{TNT}.
	
	Recall that the $\frac{1}{2}$-spinor bundle $\text{S}_{\frac{1}{2}}$  is a complex 4-dimensional vector bundle equipped with the spin connection 
	\begin{align*}\label{nabla} 
		\nabla_{e_{k}} \Psi = e_{k} \big(\Psi\big) + \frac{1}{4} g(\nabla_{e_{k}} e_{i} ,e_{j}) e^{i} \cdot e^{j} \cdot \Psi, \quad k=1,2,3,4
	\end{align*}	
	for a spinor $\Psi =(\psi_1, \psi_2, \psi_3, \psi_4)^t$.   Using  \eqref{w}, we obtain 
	\begin{equation}\label{tconnection}
		\begin{split}
			&\nabla_{e_{1}}\Psi=e_{1} \big(\Psi\big),\\
			&\nabla_{e_{2}}\Psi=e_{2} \big(\Psi\big) +\frac{1}{2\lb r^2-N^2 \rb f}\lb r e^{1} \cdot e^{2} -N e^{3} \cdot e^{4}\rb \cdot \Psi,\\
			&\nabla_{e_{3}}\Psi=e_{3} \big(\Psi\big) +\frac{1 }{2\lb r^2-N^2 \rb f}\lb r e^{1} \cdot e^{3} +N e^{2} \cdot e^{4}\rb \cdot \Psi,\\
			&\nabla_{e_{4}}\Psi=e_{4} \big(\Psi\big) -\frac{f'}{2f^2} e^{1} \cdot e^{4} \cdot \Psi +\lb \frac{N}{2\lb r^2-N^2\rb f}-\frac{f}{4N}\rb e^{2} \cdot e^{3} \cdot \Psi.
		\end{split}
	\end{equation}
	Throughout the paper, we fix the following Clifford representation
	\begin{equation*}
		\bal
		&e^{1}\mapsto	\begin{pmatrix}
			&&1&\\
			&&&1\\
			-1&&&\\
			&-1&&
		\end{pmatrix},
		&e^{2}&\mapsto\begin{pmatrix}
			&&&\I\\
			&&\I&\\
			&\I&&\\
			\I&&&
		\end{pmatrix},\\
		&e^{3}\mapsto\begin{pmatrix}
			&&&-1\\
			&&1&\\
			&-1&&\\
			1&&&
		\end{pmatrix},
		&e^{4}&\mapsto\begin{pmatrix}
			&&\I&\\
			&&&-\I\\
			\I&&&\\
			&-\I&&
		\end{pmatrix}.
		\eal
	\end{equation*}

  The  Dirac operator  is defined as 
	\begin{align*}
		D=e^{k} \cdot \nabla_{e_{k}}.
	\end{align*}
	A twistor spinor $u$ on the metric \eqref{TNT}  satisfies the twistor equation
	\begin{align}\label{twistor}
		\nabla_{e_k} u = -\frac{1}{4} e^k \cdot Du
	\end{align}
	for $k=1,2,3,4$. Then we have an integrability condition for such a spinor  (see \cite[Eq. (2.10)]{L})
	\begin{align}\label{integrability}
		\nabla_{e_{i}} Du =-R_{ii} e_{i} \cdot  u
	\end{align}
	for $i = 1, 2, 3, 4$, where $R_{ii}$ are given by \eqref{Ricci}.
	A parallel spinor $u$ is  a special twistor spinor satisfying
	\begin{align}\label{parallel}
		\nabla_{e_k} u = 0 
	\end{align}
	for $k=1,2,3,4$. Its existence implies that the metric is Ricci flat \cite[Prop. 5.12]{H2}.
	\begin{prop} 
		Nontrivial twistor spinors on the scalar flat metric \eqref{TNT} exist  only for  $f$ satisfying \eqref{TN} or \eqref{TN-N}, and they coincide with the parallel spinors.	
	\end{prop}
	\pf
	Let $u=(u_1, u_2, u_3, u_4)^t$ be a twistor spinor. From \eqref{twistor}, we obtain
	\begin{align}\label{Du}
		Du=4e^1 \cdot \nabla_{e_{1}}u = 4e^2 \cdot \nabla_{e_{2}}u = 4e^3 \cdot \nabla_{e_{3}}u = 4e^4 \cdot \nabla_{e_{4}}u.
	\end{align}
Substituting this into  \eqref{integrability} yields
	\begin{align}
		\label{integrabilityeq1} 4e^1 \cdot \nabla_{e_{1}}(e^1 \cdot \nabla_{e_{1}}u)=\frac{N^2 -C_{2}}{\lb r^2-N^2 \rb^2}u,\\
		\label{integrabilityeq4} 4e^4 \cdot \nabla_{e_{4}}(e^4 \cdot \nabla_{e_{4}}u)=\frac{N^2 -C_{2}}{\lb r^2-N^2 \rb^2}u.
	\end{align}
	The general solution of \eqref{integrabilityeq4} is given by
	\begin{align}\label{Solution}
		u_k=e^{ Q_{k+}(r)\psi}	u_{k1}(r, \theta, \phi) + e^{Q_{k-}(r)\psi}	u_{k2}(r, \theta, \phi)
	\end{align}
	for $k=1,2,3,4$, where $u_{k1}$, $u_{k2}$ are functions of $r, \theta, \phi$ and
	\begin{align*}
		Q_{1\pm}=Q_{3\pm}=&  -\frac{2\I N}{f}\lb \frac{N}{2(r^2-N^2) f} -\frac{f}{4N} \rb \\
		&\pm \sqrt{-\frac{N^2}{f^2} \lb \frac{(f')^2}{f^4} +\dfrac{N^2 -C_2}{(r^2-N^2)^{2}}\rb},\\ 
		Q_{2\pm}=Q_{4\pm}=&  \frac{2\I N}{f}\lb \frac{N}{2(r^2-N^2) f} -\frac{f}{4N} \rb \\
		&\pm \sqrt{-\frac{N^2}{f^2} \lb \frac{(f')^2}{f^4} +\dfrac{N^2 -C_2}{(r^2-N^2)^{2}}\rb}.
	\end{align*}
	Substituting \eqref{Solution} into the equation
	\begin{align}\label{Psii}
		e^1 \cdot \nabla_{e_{1}}u-	e^4 \cdot \nabla_{e_{4}}u=0,
	\end{align}
	we obtain  the constraints
	\begin{align*}
		u_{12} = u_{21} = u_{32} = u_{41} = 0,
	\end{align*}
    and
	\begin{align*}
	C_2 = \frac{(C_1)^2}{4}.	
	\end{align*} 
  Then \eqref{Psii} reduces to
	\begin{align}\label{partialru11}
		\begin{split}
			\partial_r u_{i} + \frac{C_1 - 2N}{2(r+N)(2r +C_1)}  u_{i}=0,\quad i=1,2,\\
			\partial_r u_{j} +  \frac{C_1 + 2N}{2(r-N)(2r +C_1)} u_{j}=0,\quad j=3,4,
		\end{split}
	\end{align}
	and \eqref{integrabilityeq1} becomes
	\begin{align}\label{partialrru112}
		\begin{split}
			\partial_{r} \partial_{r} u_{k} - \frac{C_1 r +2N^2}{(r^2-N^2)(2r + C_1)} \partial_{r} u_{k} +\frac{  N^2 -\frac{(C_1)^2}{4} }{(r^2-N^2)(2r+C_1)^2} u_{k}=0 
		\end{split}
	\end{align} 
for $k=1,2,3,4$.	Combining  \eqref{partialru11}, \eqref{partialrru112} and the partial derivative with respect to $r$ of \eqref{partialru11}, we find that two distinct cases occur:
	\begin{align*}
		C_1=2N, \quad C_2=N^2,
	\end{align*}
	\begin{align*}
		\partial_r u_{11}=0, \quad \partial_r u_{22}=0, \quad u_{31}=0, \quad u_{42}=0,
	\end{align*}
	or
	\begin{align*}
		C_1=-2N, \quad C_2=N^2,
	\end{align*}
	\begin{align*}
		u_{11}=0, \quad  u_{22}=0, \quad \partial_r u_{31}=0, \quad  \partial_r u_{42}=0.
	\end{align*}
	In both cases, we have $\nabla_{e_{1}}u =0$, so that  \eqref{Du} gives
	\begin{align*}
		\nabla_{e_{1}} u = \nabla_{e_{2}} u = \nabla_{e_{3}} u = \nabla_{e_{4}} u =0.
	\end{align*}
	This completes the proof. \qed

Note that in \cite{W1}, Wang classified the holonomy groups of manifolds admitting parallel spinors and showed that the space of parallel spinors is complex $2$-dimensional for both the Taub-NUT metric and its negative NUT charge  counterpart.
We verify this result by directly solving \eqref{parallel}. The parallel spinor for Taub-NUT metric is given by
	\begin{align}\label{TNp}
		u= C_3e^{-\frac{\I}{2}\phi}\begin{pmatrix}
			0\\
			0\\
			e^{\frac{\I}{2}\psi}\sin\frac{\theta}{2}\\
			e^{-\frac{\I}{2}\psi}\cos\frac{\theta}{2}
		\end{pmatrix}+ C_4e^{\frac{\I}{2}\phi}\begin{pmatrix}
			0\\
			0\\
			e^{\frac{\I}{2}\psi}\cos\frac{\theta}{2}\\
			-e^{-\frac{\I}{2}\psi}\sin\frac{\theta}{2}
		\end{pmatrix},
	\end{align} 
while for the Taub-NUT metric with negative NUT charge, it takes the form
	\begin{align}\label{TN-Np}
		u= C_3e^{-\frac{\I}{2}\phi}\begin{pmatrix}
			e^{\frac{\I}{2}\psi}\sin\frac{\theta}{2}\\
			- e^{-\frac{\I}{2}\psi}\cos\frac{\theta}{2}\\
			0\\
			0
		\end{pmatrix}+ C_4e^{\frac{\I}{2}\phi}\begin{pmatrix}
			e^{\frac{\I}{2}\psi}\cos\frac{\theta}{2}\\
			e^{-\frac{\I}{2}\psi}\sin\frac{\theta}{2}\\
			0\\
			0
		\end{pmatrix},
	\end{align}
	where $C_3, C_4$ are complex constants.

	On a $4$-dimensional Riemannian spin manifold $M$ equipped with the scalar flat Taub-NUT type  metric \eqref{TNT}, the Dirac equation is
	\begin{align}\label{Dirac}
		D\Psi=e^{k} \cdot \nabla_{e_{k}} \Psi = \lambda \Psi.
	\end{align}
	Since $M$ is noncompact, $\lambda$ is generally a complex number, and there are point spectrum, essential spectrum, discrete spectrum and  continuous spectrum (cf. \cite[Def. 7.1.2]{G}).
	
	We next introduce the Rarita-Schwinger equation on the $\frac{3}{2}$-spinor bundle  $\mathrm{S}_{\frac{3}{2}}$ over  $M$. Starting from the tensor bundle 
	\begin{align*}
		\mathrm{S}_{\frac{1}{2}} \otimes T^{*}M = \left\{\Psi_i \otimes e^i \;\middle| \;  \Psi_i \in \mathrm{S}_{\frac{1}{2}},\ i=1,2,3,4 \right\},
	\end{align*}
	we define the complex scalar multiplication by
	\begin{align*}
		z \lb \Psi_i \otimes e^i \rb = \lb z \Psi_i  \rb \otimes e^i, \quad \forall\,z \in \mathbb{C}.
	\end{align*}
	The metric \eqref{TNT} induces a Riemannian inner product on $T^*M$, which we still denote by $g$. The Hermitian metric on $\mathrm{S}_{\frac{1}{2}} \otimes T^*M $ is given by
	\begin{align*}
		\lb \Psi_i \otimes e^i, \hat{\Psi}_j \otimes e^j  \rb  = \lb \Psi_i,  \hat{\Psi}_j \rb g\lb e^i, e^j \rb  
		= \sum\limits_{i=1}^{4}\lb\Psi_i, \hat{\Psi}_i \rb,
	\end{align*}
	where $\lb  \cdot, \cdot \rb$ on $\mathrm{S}_{\frac{1}{2}}$  is the  natural Hermitian inner product. The Clifford multiplication on $\mathrm{S}_{\frac{1}{2}} \otimes T^{*} M $ is defined as
	\begin{align*}
		\alpha \cdot \lb \Psi_i \otimes e^i \rb =   \lb \alpha  \cdot \Psi_i \rb \otimes e^i, \quad \forall\,\alpha \in T^*M,
	\end{align*} 
	and the covariant connection  is given by
	\begin{align*}
		\nabla_{X}(\Psi_{i} \otimes e^{i}) =\nabla_{X} \Psi_{i} \otimes e^{i}+\Psi_{i}\otimes \nabla_{X} e^{i}, \quad \forall\,X \in TM.
	\end{align*}
	We have the following identities:
	\begin{align*}
		\alpha \cdot \beta \cdot(\Psi_{i} \otimes e^{i})+	 \beta \cdot \alpha\cdot \lb \Psi_{i} \otimes e^{i}\rb =& -2g(\alpha, \beta) \Psi_{i} \otimes e^{i},\\
		\lb \alpha \cdot \lb \Psi_{i} \otimes e^{i} \rb , \hat{\Psi}_j \otimes e^j  \rb = & -\lb  \Psi_{i} \otimes e^{i},  \alpha \cdot \big( \hat{\Psi}_j \otimes e^j \big) \rb,\\
		\nabla_{X} \lb \alpha \cdot \lb  \Psi_i \otimes e^i \rb \rb = & \lb  \nabla_{X}  \alpha \rb \cdot \lb  \Psi_i \otimes e^i  \rb  +  \alpha \cdot  \nabla_{X}\lb  \Psi_i \otimes e^i \rb,\\
		X\lb  \Psi_{i} \otimes e^{i},  \hat{\Psi}_j \otimes e^j \rb = & \lb  \nabla_{X} \lb \Psi_{i} \otimes e^{i}\rb,  \hat{\Psi}_j \otimes e^j \rb \\
		& +\lb  \Psi_{i} \otimes e^{i}, \nabla_{X}\big( \hat{\Psi}_j \otimes e^j \big)  \rb.
	\end{align*}
	
	The twisted Dirac operator 
	\begin{align*}
		D_{TM} = e^{j} \cdot \nabla_{e_j}
	\end{align*}
	is defined  by 
	\begin{align*}
		D_{TM}(\Psi_{i} \otimes e^{i})=D\Psi_{i} \otimes  e^{i}+ e^j \cdot \Psi_{i} \otimes \nabla_{e_j} e^{i}.	
	\end{align*}

	We now consider the $\frac{3}{2}$-spinor bundle $\mathrm{S}_{\frac{3}{2}}$ over $M$. The projection from $\mathrm{S}_{\frac{1}{2}} \otimes T^{\ast}M$ to $\mathrm{S}_{\frac{3}{2}}$ is 
	\begin{align*}
		\Pi( \Psi_i \otimes e^i) = \Psi_i \otimes e^i+\frac{1}{4}e_i\cdot(e^j\cdot\Psi_j)\otimes e^i.
	\end{align*}
	If we take 
	\begin{align*}
		-\frac{1}{4}e_i\cdot(e^j\cdot\Psi_j)\otimes e^i= \Pi(\Psi_i \otimes e^i),
	\end{align*}
	it follows that $\Psi_i = 0$ for $i=1,2,3,4$. This leads to a direct sum decomposition
	\begin{align*}
		\mathrm{S}_{\frac{1}{2}} \otimes T^{\ast}M & \longrightarrow {\mathrm{S}_{\frac{3}{2}}}^{\perp} \oplus \mathrm{S}_{\frac{3}{2}} \\
		\Psi_i \otimes e^i & \longmapsto\lb-\frac{1}{4}e_i\cdot(e^j\cdot\Psi_j)\otimes e^i,\,\Pi( \Psi_i \otimes e^i) \rb.
	\end{align*}
	Define a linear map  $\mu$ from ${\mathrm{S}_{\frac{3}{2}}}^{\perp}$ to $\mathrm{S}_{\frac{1}{2}}$ by 
	\begin{align*}
		\mu\lb -\frac{1}{4}e_i\cdot(e^j\cdot\Psi_j)\otimes e^i\rb=e^j\cdot\Psi_j.
	\end{align*}
	Since the kernel of $\mu$ is trivial, $\mu$ is injective.  To show surjectivity, given any $\Psi \in \mathrm{S}_{\frac{1}{2}}$,
	take $\Psi_j = -\frac{1}{4}e_j \cdot\Psi$, $j=1,2,3,4$. Then 
	\begin{align*}
		\mu\lb -\frac{1}{4}e_i\cdot(e^j\cdot\Psi_j)\otimes e^i\rb=\Psi.
	\end{align*}
	Thus $\mu$ is bijective,  and we have $	{\mathrm{S}_{\frac{3}{2}}}^{\perp} \cong \mathrm{S}_{\frac{1}{2}}$. Consequently,  
	\begin{align*}
		\mathrm{S}_{\frac{1}{2}} \otimes T^{\ast}M \cong \mathrm{S}_{\frac{1}{2}} \oplus \mathrm{S}_{\frac{3}{2}}.
	\end{align*}
	In this paper, we always consider the $\frac{3}{2}$-spinor bundle as
	\begin{align*}
		\mathrm{S}_{\frac{3}{2}} = \left\{\Psi_i\otimes e^i \in 	\mathrm{S}_{\frac{1}{2}} \otimes T^{*}M \;\middle| \; e^i \cdot \Psi_i=0 \right\}.
	\end{align*}
	
	The Rarita-Schwinger operator is defined as 
	\begin{align*}
		Q=\Pi\circ D_{TM} \big|_{ \mathrm{S}_{\frac{3}{2}}}: \text{S}_{\frac{3}{2}} \longrightarrow \text{S}_{\frac{3}{2}},
	\end{align*}
	and the Rarita-Schwinger equation is given by
	\begin{align}\label{RS}
		Q (\Psi_{i} \otimes e^{i}) = \lambda \Psi_{i} \otimes e^{i}.
	\end{align}
Specifically, a Rarita-Schwinger field  satisfies 
	\begin{align}\label{rsfeild}	 
		Q (\Psi_{i} \otimes e^{i}) =0,
	\end{align}
and
	\begin{align}\label{divergence}	
		\sum_{i=1}^{4}\nabla_{e_i}(\Psi_{i} \otimes e^{i}) (e_i) = 0.
\end{align}

	\mysection{Massless solutions}\ls
	In this section, we study the harmonic spinors and  Rarita-Schwinger fields on the metric \eqref{TNT} with $f$ given by \eqref{TN} or \eqref{TN-N}, and analyze their $L^2$ integrability.
	
	In  \cite{AE}, A\c{c}\i k and Ertem constructed harmonic spinors on 4-dimensional Ricci flat metrics using a parallel spinor $u$ via
	\begin{align*}
			\Psi = \d \varphi \cdot u, 
	\end{align*}
    or
    \begin{align*}
		\Psi = F \cdot u,
    \end{align*}
   where $\varphi$ is a harmonic function satisfying 
   \begin{align*}
	\sum_{i=1}^{4} (\nabla_{e_i}\d \varphi) (e_i)= 0,
   \end{align*}
	and $F$ is a  Maxwell field obeying
   \begin{align}\label{Maxwell}
   	\begin{split}
   		\d F = 0,\\
        \sum_{i=1}^{4} (\nabla_{e_i}F)(e_i)=0.
      \end{split}
   \end{align}
	They also constructed Rarita-Schwinger fields as
    \begin{align*}
		\sigma=F\cdot {e_{i}}\cdot u \otimes e^i,
   \end{align*}
or
   \begin{align*}
	\sigma=\nabla_{e_i}F \cdot u\otimes e^i. 
   \end{align*}
We now apply this framework to the metric \eqref{TNT}.  For  $f$ given by \eqref{TN} and \eqref{TN-N}, we take the harmonic function
	\begin{align*}
	\varphi=-\frac{1}{r \mp N},
  \end{align*}
respectively.
The canonical solution  of \eqref{Maxwell} for both cases is the self-dual 2-form
	\begin{align}\label{F}
		F=\frac{1}{(r+N)^2}\lb e^1 \wedge e^4+  e^2 \wedge e^3\rb.
	\end{align}
Moreover, for the parallel spinor $u$ given by \eqref{TNp} and \eqref{TN-Np}, we find 
\begin{align*}
	\lvert u \rvert^2 = \lv C_3\rv^2 +\lv C_4\rv^2,
\end{align*}
so we can set  $\lvert u \rvert=1$ without loss of generality.
	\begin{thm}
		For the metric \eqref{TNT} with $f$ given by \eqref{TN}, the harmonic spinor
		\begin{align}\label{Harmonic+}
			\Psi=\frac{e^1 \cdot u }{(r+N)^{\frac{1}{2}}(r-N)^{\frac{3}{2}}}
		\end{align}
	 is not in $L^p$ for $0 < p < \infty$, where $u$ is provided by \eqref{TNp}. 
	\end{thm}
	\pf
	Let $\d\mu$ be the volume element of the metric \eqref{TNT}. There exist a constant $C' > 0$ and a sufficiently large $r_1>N$ such that for $r \geq r_{1}$ and  $0 < p \leq \frac{3}{2}$, 
	\begin{align*}
		\int_{\mathcal{D}} \int_{ r_1}^{\infty} \lv \Psi \rv^p \d\mu&=32N \pi^2 \int_{r_1}^{\infty}\frac{\lv u\rv^p (r^2 -N^2)}{(r+N)^{\frac{p}{2}}(r-N)^{\frac{3p}{2}}}\d r\\
		&>C'\int_{r_1}^{\infty}\frac{1}{r^{2p-2}}\d r=\infty.
	\end{align*}
	In contrast, for $N \leq r < r_{1}$ and $\frac{4}{3} \leq p < \infty$, we obtain
	\begin{align*}
		\int_{\mathcal{D}} \int_{ N}^{r_1}\lv \Psi\rv^p \d\mu>C''\int_{N}^{r_1}\frac{1}{(r-N)^{\frac{3p}{2}-1}}\d r=\infty.
	\end{align*}
	Therefore, $\Psi$ given by \eqref{Harmonic+} is not in $L^p$ for any $0 < p < \infty$. \qed

	\begin{rmk}
		 The spinor $\Psi = F \cdot u$ vanishes identically for $u$ given by \eqref{TNp} and $F$ from \eqref{F}.
	\end{rmk}

	\begin{thm}
	For the metric \eqref{TNT} with $f$ given by \eqref{TN-N},  there exist two $L^2$ harmonic spinors  
		\begin{align}\label{Harmonic-} 
			\Psi&=\frac{e^1 \cdot u }{(r-N)^{\frac{1}{2}} (r+N)^{\frac{3}{2}}},
		\end{align}
		and
		\begin{align}\label{DMaxwell-}
			\Psi&= \frac{1}{(r+N)^{2}} \lb e^1 \cdot e^4 + e^2 \cdot e^3 \rb \cdot u,
		\end{align}
		where $u$ is provided by \eqref{TN-Np}.
	\end{thm}
	\pf  For $\Psi$ given by \eqref{Harmonic-}, we obtain
	\begin{align*}
		\int_{\mathcal{D}} \int_{N}^{\infty} \lv \Psi\rv^2 \d\mu =32N\pi^2 \int_{N}^{\infty} \frac{1}{(r+N)^2}  \d r = 16\pi^2 < \infty.
	\end{align*}
	For $\Psi$ given by \eqref{DMaxwell-}, we have
	\begin{align*}
		\lv \Psi\rv^2 &=\frac{1}{(r+N)^4}
		\lb \lb e^1\cdot e^4+e^2\cdot e^3 \rb \cdot u,  
		\lb e^1\cdot e^4+e^2\cdot e^3 \rb \cdot u\rb  \\
		&=  \frac{4 \lvert u \rvert^2}{(r+N)^4}.
	\end{align*}
Integrating it gives
	\begin{align*}
			\int_{\mathcal{D}} \int_{N}^{\infty} \lv \Psi\rv^2 \d\mu =32 \pi^2 < \infty.
	\end{align*}
Therefore, both harmonic spinors \eqref{Harmonic-} and \eqref{DMaxwell-} are $L^2$.
  \qed
	
	\begin{thm}
		For the metric \eqref{TNT} with $f$ given by \eqref{TN} ,  an $L^2$ Rarita-Schwinger field is 
		\begin{align}\label{RSfieldTN}
			\sigma = \frac{1}{(r+N)^2}\lb e^1 \cdot e^4 + e^2 \cdot e^3\rb  \cdot e_i \cdot u \otimes e^i,
		\end{align}
     	where $u$ is provided by \eqref{TNp}. 
     	In the case with $f$ given by \eqref{TN-N},    an $L^2$ Rarita-Schwinger field is
		\begin{align}\label{RSfieldTN-}
			\begin{split}
		\sigma=&\frac{1}{(r-N)^{\frac{1}{2}}(r+N)^\frac{5}{2} } \biggl( -2 \lb e^1\cdot e^4+ e^2\cdot e^3\rb \cdot u\otimes e^1 \\
		       &+\lb e^2 \cdot e^4 - e^1 \cdot e^3 \rb \cdot u \otimes e^2 + \lb e^1 \cdot e^2 + e^3 \cdot e^4 \rb \cdot u \otimes e^3 \biggr),
			\end{split}
		\end{align}
    	where $u$ is provided by \eqref{TN-Np}.
	\end{thm}
	\pf For $\sigma$ given by \eqref{RSfieldTN}, we obtain
	\begin{align*}
		\lv \sigma \rv^2=\frac{1}{(r+N)^4}\lb E\cdot u, E\cdot u\rb 
		=\frac{16\lv u \rv^2}{(r+N)^4},
	\end{align*}
	where 
	\begin{align*}
		E=\lb e^1\cdot e^4+ e^2\cdot e^3\rb\cdot \lb e_1+e_2+e_3+e_4 \rb.
	\end{align*}
	Thus,  we have
	\begin{align*}
		\int_{\mathcal{D}} \int_{N}^{\infty} \lv \sigma\rv^2 \d\mu =128 \pi^2 < \infty.
	\end{align*} 
	For $\sigma$ from \eqref{RSfieldTN-}, we obtain
	\begin{align*}
		\lv \sigma \rv^2 &=\frac{6}{(r-N)(r+N)^5} \lb (1-e^1\cdot e^2\cdot e^3\cdot e^4) \cdot u,  (1-e^1\cdot e^2\cdot e^3\cdot e^4) \cdot u  \rb \\
		  &=\frac{24\lv u\rv ^2}{(r-N)(r+N)^5}.
	\end{align*}
It follows that
	\begin{align*}
		\int_{\mathcal{D}} \int_{N}^{\infty} \lv \sigma\rv^2 \d\mu
		=768N \pi^2\int_{N}^{\infty} \frac{1}{(r+N)^4}\d r=\frac{32\pi^2}{N^2}<\infty.
	\end{align*}
 Therefore, both Rarita-Schwinger fields \eqref{RSfieldTN} and \eqref{RSfieldTN-} are $L^2$.  \qed

	\mysection{Separation of the Dirac equation}\ls
	In this section, we separate the Dirac equation \eqref{Dirac} on the metric \eqref{TNT} into angular and radial equations by the spinor ansatz
	\begin{align}\label{TNTform}
		\begin{pmatrix}
			\psi_{1}\\
			\psi_{2}\\
			\psi_{3}\\
			\psi_{4}
		\end{pmatrix}=e^{\I\lb m+\frac{1}{2}\rb\phi}\begin{pmatrix}
			e^{\frac{\I}{2} \lb m_1 +\frac{1}{2}\rb \psi}\Phi_{1}(r)J_{+}(\theta)\\
			e^{\frac{\I}{2} \lb m_2 +\frac{1}{2}\rb\psi}\Phi_{2}(r)J_{-}(\theta)\\
			e^{\frac{\I}{2} \lb m_1+\frac{1}{2}\rb \psi}\Phi_{3}(r)J_{+}(\theta)\\
			e^{\frac{\I}{2} \lb m_2 +\frac{1}{2}\rb \psi}\Phi_{4}(r)J_{-}(\theta)
		\end{pmatrix},	
	\end{align}
	where $m, m_1, m_2$ are integers. 
	
	Using the spin connection \eqref{tconnection}, the Dirac equation on the metric \eqref{TNT} can be written  as
	\beQ\label{Diracequation}	
	\frac{1}{\sqrt{r^2-N^2}}\begin{pmatrix}
		0&0& \mathcal{D}_{1+}& 	e^{\I \psi}\mathcal{L}_{+}\\[3pt]
		0&0& 	e^{-\I \psi}\mathcal{L}_{-}& \mathcal{D}_{1-}\\[3pt]
		-\mathcal{D}_{2-} & e^{\I \psi}\mathcal{L}_{ +} &0&0\\[3pt]
		e^{-\I \psi}\mathcal{L}_{-}& -\mathcal{D}_{2+}&0&0
	\end{pmatrix}
	\begin{pmatrix}
		\psi_{1}\\[3pt] \psi_{2}\\[3pt] \psi_{3}\\[3pt] \psi_{4}
	\end{pmatrix}
	=\lambda\begin{pmatrix}
		\psi_{1}\\[3pt] \psi_{2}\\[3pt] \psi_{3}\\[3pt]\psi_{4}
	\end{pmatrix},
	\eeQ
	where
	\begin{align*}
		\mathcal{L}_{\pm} = & \pm \partial_{\theta} - \frac{\I}{\sin\theta}\partial _{\phi} + \frac{\I\cos\theta}{\sin\theta}\partial _{\psi},\\
		\mathcal{D}_{s\pm} =&\sqrt{r^2-N^2} \lb \frac{1}{f} \partial_{r} \pm \frac{\I f}{2N}\partial_{\psi} + h_s \rb,\\
		h_s=&(-1)^s \lb  \frac{N}{2(r^2-N^2)f}-\frac{f}{4N}\rb + \frac{1}{(r+ (-1)^s N)f}-\frac{f'}{2f^2}
	\end{align*}
	for $s = 1, 2$. The separation of the above equations via \eqref{TNTform} yields the angular  equations
	\begin{align}\label{theta}
		\begin{split}
			\frac{1}{J_{-}} \lb -\partial_{\theta} +\frac{m +\frac{1}{2}}{\sin\theta} -\lb \frac{m_1}{2} +\frac{1}{4}\rb \frac{\cos\theta}{\sin\theta} \rb J_{+} &=\eta e^{
				\I \lb \frac{1}{2}(m_2 -m_1) +1 \rb \psi},\\
			\frac{1}{J_{+}}\lb \partial_{\theta} +\frac{m +\frac{1}{2}}{\sin\theta} -\lb \frac{m_2}{2} +\frac{1}{4}\rb \frac{\cos\theta}{\sin\theta}\rb J_{-} &=\eta e^{
				\I \lb\frac{1}{2}(m_1 -m_2) -1\rb \psi},
		\end{split}
	\end{align}
  and the radial equations
	\begin{align}\label{r}
		\begin{split}	
		\partial_r \Phi_{1}=&\lb f h_2 -\frac{f^2}{4N} \lb m_1 +\frac{1}{2}\rb  \rb \Phi_{1} +\frac{\eta  }{\sqrt{r^2 +C_1 r +C_2}} \Phi_{2} -\lambda f \Phi_{3},\\	
		   \partial_r \Phi_{2}=&\lb f h_2 +\frac{f^2}{4N} \lb m_2 +\frac{1}{2}\rb  \rb \Phi_{2} +\frac{\eta  }{\sqrt{r^2 +C_1 r +C_2}} \Phi_{1} -\lambda f \Phi_{4},\\	
	        	\partial_r \Phi_{3}=&\lb -f h_1  +  \frac{f^2}{4N} \lb m_1 +\frac{1}{2}\rb\rb \Phi_{3} -\frac{\eta }{\sqrt{r^2 +C_1 r +C_2}} \Phi_{4} +\lambda f \Phi_{1},\\		
		            	\partial_r \Phi_{4}=&\lb -f h_1 - \frac{f^2}{4N} \lb m_2 +\frac{1}{2}\rb\rb \Phi_{4} -\frac{\eta }{\sqrt{r^2 +C_1 r +C_2}} \Phi_{3} +\lambda f \Phi_{2},
		\end{split}
	\end{align}	
	where $\eta$ is a complex constant.	 As the left hand side of \eqref{theta} depends only on $\theta$, the right hand side must be constant. Therefore two cases occur 
	\begin{align}\label{2cases}
		\mbox{(i)}\,\, \eta \neq 0,\quad  m_1-m_2=2,  \quad \mbox{(ii)}\,\, \eta = 0.
	\end{align}
	
	The angular equations \eqref{theta} are preserved no matter whether the metric \eqref{TNT} is Ricci flat or scalar flat. In \cite{SU}, Sucu and \"Unal solved \eqref{theta} for the case (i) 
	\begin{align*}
		J_{+}=&\lb \sin\frac{\theta}{2} \rb^{m-\frac{m_1}{2} +\frac{1}{4}}
		\lb \cos\frac{\theta}{2} \rb^{\gamma-1} F\lb \alpha,\beta; \gamma; \lb \cos\frac{\theta}{2} \rb^2\rb,\\
		J_{-}=&-\frac{\eta}{\gamma}
		\lb \sin\frac{\theta}{2} \rb^{m-\frac{m_1}{2} +\frac{5}{4}}\lb \cos\frac{\theta}{2} \rb^{\gamma}  
		F \lb \alpha+1, \beta+1; \gamma+1; \lb \cos\frac{\theta}{2}\rb^2 \rb,
	\end{align*}
	where  $F$ is the hypergeometric function (cf. \cite[Chap. 5]{K}), and
	\begin{align*}
		\alpha&=\frac{1}{4}-\frac{m_1}{2} +\sqrt{\lb \frac{1}{4}-\frac{m_1}{2}\rb^2+\eta^2},\\
		\beta&=\frac{1}{4}-\frac{m_1}{2}-\sqrt{\lb \frac{1}{4}-\frac{m_1}{2}\rb^2+\eta^2},\\
		\gamma&=\frac{1}{4}-\frac{m_1}{2}-m.
	\end{align*}
We find that $\begin{pmatrix} J_{+}\\	J_{-} \end{pmatrix}$  is singular at $\theta=\pi$ for $ m >  -\frac{m_1}{2}-\frac{3}{4} $, and  at $\theta=0$ for any integer $m$.
In the Cartesian coordinates defined by
\begin{align*}
	x_{1}=r \cos \frac{\theta}{2} \cos \frac{\psi+\phi}{2}, &\quad x_{2}=r \cos \frac{\theta}{2} \sin \frac{\psi+\phi}{2}, \\
	x_{3}=r \sin \frac{\theta}{2} \cos \frac{\psi-\phi}{2}, &\quad  x_{4}=r \sin \frac{\theta}{2} \sin \frac{\psi-\phi}{2},
\end{align*}
these singularities lie on the $x_{3}x_{4}$-plane when $ m >  -\frac{m_1}{2}-\frac{3}{4}$, and on the $x_{1}x_{2}$-plane where $x_3=x_4=0$.

The angular equations \eqref{theta} for the case (ii) are solved as follows.	
	
\begin{thm}\label{prop-angularpart}
		Solutions of \eqref{theta} for $\eta=0$ are	given by
		\begin{align}\label{angular0}
			\begin{split}
				J_{+}&=\lb\sin\frac{\theta}{2}\rb^{
					m-\frac{m_1}{2}+\frac{1}{4}}
				\lb\cos\frac{\theta}{2}\rb^{-m-\frac{m_1}{2}-\frac{3}{4}},\\
				J_{-}&=\lb\sin\frac{\theta}{2}\rb^{
					\frac{m_2}{2}-m-\frac{1}{4}}\lb\cos\frac{\theta}{2}\rb^{m+\frac{m_2}{2}+\frac{3}{4}},
			\end{split}
		\end{align} 	
		which are regular if
		\begin{align}\label{Dmgeq0}
			m\geq 0,\quad 
			m_1 \leq  -2m-\frac{3}{2},\quad
			m_2 \geq  2 m+\frac{1}{2},
		\end{align}
		or 
		\begin{align}\label{Dm<0}
			m < 0, \quad m_1 \leq  2m+\frac{1}{2} ,\quad
			m_2 \geq  -2m-\frac{3}{2}.	
		\end{align}
	\end{thm}	
	\pf   Setting $\eta=0$ in \eqref{theta}, we obtain	
	\begin{align*}
		\partial_{\theta}J_{+}&=\lb \frac{m+\frac{1}{2}}{\sin\theta}  -\lb \frac{m_1}{2}+\frac{1}{4}\rb\frac{\cos\theta}{\sin\theta} \rb J_{+}, \\ \partial_{\theta}J_{-}&=\lb-\frac{m+\frac{1}{2}}{\sin\theta}+\lb \frac{m_2}{2}+\frac{1}{4}\rb\frac{\cos\theta}{\sin\theta}\rb J_{-}.
	\end{align*}
	Then \eqref{angular0} follows by direct integration.  The regularity follows if the exponents of $\sin{\frac{\theta}{2}} $ and $\cos \frac{\theta}{2}$ are nonnegative.\qed

	Next we solve the radial equations \eqref{r} for $\lambda=0$. 	 Denote
	\begin{align*}
		r_0:= \frac{ -C_1+ \sqrt{(C_1)^2- 4C_2} }{2}< N.
	\end{align*}

\begin{thm}
	Let $g$ be the scalar flat Taub-NUT type metric \eqref{TNT} with $f$ given by \eqref{TNTf} and $C_2 > -N^2 -N C_1$. Suppose
	\begin{align*}
		\lambda=\eta=0.
	\end{align*}
	Nonzero  solutions of the radial equations \eqref{r} on $r>N$ are
	\begin{align}\label{harmoniceta0}
		\begin{split}
			\Phi_{3}&=\frac{(r-r_{0})^{\frac{\lb 2m_1-1\rb
						\lb r_{0}^2-N^2\rb}{8N(2r_0+C_1)}-\frac{1}{4}}
				e^{\frac{\lb 2m_1-1\rb r}{8N}}}
			{(r-N)^{\frac{1}{2}}(r+r_0+C_1)^{-\frac{\lb 2m_1-1\rb\lb N^2-(r_0+C_1)^2\rb}{8N(2r_0+C_1)}
					+\frac{1}{4}}},\\
			\Phi_{4}&=\frac{(r-r_{0})^{-\frac{(2m_2+3)
						\lb r_0^2-N^2\rb}{8N(2r_0+C_1)}    -\frac{1}{4}}
				e^{-\frac{\lb 2m_2+3\rb r}{8N}}}
			{(r-N)^{\frac{1}{2}}(r+r_0+C_1)^{\frac{(2m_2+3)\lb N^2-(r_0+C_1)^2\rb}{8N(2r_0+C_1)}
					+\frac{1}{4}}}.
		\end{split}
	\end{align}
	Solutions of the angular equations \eqref{theta} are given by \eqref{angular0}. Moreover, under the conditions \eqref{Dmgeq0} or \eqref{Dm<0}, these solutions are $L^2$ integrable. 
\end{thm}
\pf Solving \eqref{r} with $\lambda=\eta=0$ yields the solutions \eqref{harmoniceta0}. We observe that near $r=N$,
\begin{align*}
	(r^2 - N^2) ( \lv \Phi_{3} \rv^2 + \lv \Phi_{4} \rv^2 )
\end{align*}
is bounded.  For sufficiently large $r$, we have the asymptotic behavior 
 \begin{align*}	
	(r^2 - N^2) \lv \Phi_{3}\rv^2 & \sim r^{-\frac{(2m_1-1)C_1}{4N}}  e^{\frac{\lb 2m_1 - 1\rb r}{4N}},\\
	(r^2 - N^2) \lv \Phi_{4}  \rv^2 & \sim r^{\frac{(2m_2+3)C_1}{4N}} 
	e^{-\frac{\lb 2m_2 + 3\rb r}{4N}}.
\end{align*}
Under the conditions \eqref{Dmgeq0} or \eqref{Dm<0} that ensure $m_1\leq 0$ and  $m_2 \geq -1$ in \eqref{TNTform}, these functions decay to zero as  $r \rightarrow \infty$, and $J_{\pm}$ from \eqref{angular0} are regular. Thus, there exists a constant $C'$ such that
\begin{align*}
	\int_{\mathcal{D}} \int_{N}^{\infty} \lv \Psi\rv^2 \d\mu < C'\int_{ N}^{\infty}  (r^2-N^2) \lb \lv \Phi_{3}\rv^2 + \lv \Phi_{4}  \rv^2 \rb \d r < \infty.
\end{align*}   
This completes the proof.    \qed

   In the following  we study the  radial equations  \eqref{r} for $\lambda \neq 0$ and express their solutions in terms of  Kummer functions.	
	The	Kummer  equation  is defined as 
	\begin{align}\label{Kummer}
		zw''(z)+\lb \gamma-z\rb w'(z)-\alpha w(z)=0
	\end{align}
	with  complex numbers  $ \alpha, \gamma$, which has a
	regular singular point at $z=0$ and an irregular singular point at $z=\infty$ (cf. \cite[Chap. 7]{K}).
	Its solution can be represented by the Kummer function
	\begin{align*}
		w(z)=\,_1F_1\lb  \alpha; \gamma; z  \rb
		=\sum_{n=0}^{\infty}\frac{(\alpha, n)}{(\gamma,n)}\frac{z^n}{n!},
	\end{align*}
	where $(\alpha, n)$ is the Pochhammer symbol, and $\gamma$  is not a nonpositive integer.
	This power series converges everywhere in the finite complex plane, i.e., $\lv z\rv<\infty$, and 
	\begin{align*}
		_1F_1\lb  \alpha; \gamma; 0  \rb =1.
	\end{align*}
	The derivative of this function is given by
	\begin{align}\label{dKummer}
		\frac{\d}{\d z}\,_1F_1\lb\alpha;\gamma;z \rb=\frac{\alpha}{\gamma}\,_1F_1\lb\alpha+1;\gamma+1;z \rb.
	\end{align}
	We fix the following branches throughout this paper. For any real numbers $x, y$, and for $k=0, 1$, we define 
	\begin{align*}
		\sqrt{x^2-\lambda^2y^2} = \left\{\begin{array}{ll}
			\sqrt{\lv x^2-\lambda^2y^2\rv}\, e^{\frac{\I}{2}\arccos\lb \frac{x^2-(a^2-b^2)y^2}{\lv x^2-\lambda^2y^2\rv}\rb+\I k\pi},& ab\leq 0,\\
			\sqrt{\lv x^2-\lambda^2y^2\rv}\, e^{-\frac{\I}{2}\arccos\lb \frac{x^2-(a^2-b^2)y^2}{\lv x^2-\lambda^2y^2\rv}\rb+\I k\pi}, & ab> 0.
		\end{array}\right.
	\end{align*}
	
\begin{thm}\label{DiracTN-Nthm4.5}
	Let $g$ be the scalar flat Taub-NUT type metric \eqref{TNT} with $f$ given by \eqref{TN-N}. Suppose
	\begin{align*}
		\lambda=a+\I b\neq0, \quad \eta=0.
	\end{align*}
	Solutions of the radial equations \eqref{r} on $r> N$ are
	\begin{align*}
			\Phi_{1}=&\frac{(r+N)^{\frac{m_1}{2}-\frac{5}{4}}}{\sqrt{e^{z_1}}}
			\,_1F_1\lb  \alpha_1;
			m_1-\frac{1}{2}  ; z_1 \rb,\\
			\Phi_{2}=&\frac{(r+N)^{-\frac{m_2}{2}-\frac{7}{4}}}{\sqrt{e^{z_2}}}
			\,_1F_1\lb  \alpha_2;
			-m_2-\frac{3}{2}  ; z_2 \rb,\\
			\Phi_{3}=&\frac{(r+N)^{\frac{m_1}{2}-\frac{3}{4}}}{8N\lambda \sqrt{(r-N)e^{z_1}}}\lb (1-2m_1+\epsilon_1)\,_1F_1\lb  \alpha_1;
			m_1-\frac{1}{2}  ; z_1 \rb\right.\\&\left.-\frac{\epsilon_1^2+\epsilon_1(1-2m_1)+32N^2\lambda^2}{1-2m_1}\,_1F_1\lb  \alpha_1+1;
			m_1+\frac{1}{2}  ;z_1\rb\rb,\\
			\Phi_{4}=&\frac{(r+N)^{-\frac{m_2}{2}-\frac{5}{4}}}{8N\lambda \sqrt{(r-N)e^{z_2}} }\lb
			(2m_2+3+\epsilon_2)	\,_1F_1\lb  \alpha_2;
			-m_2-\frac{3}{2}  ; z_2\rb\right.\\&\left.-\frac{\epsilon_2^2+\epsilon_2(2m_2+3)+32N^2\lambda^2}{2m_2+3} 	
			\,_1F_1\lb  \alpha_2+1;
			-m_2-\frac{1}{2}  ; z_2\rb\rb,
	\end{align*}
	where
	\begin{align}\label{z}
		&z_s (r)=\frac{\epsilon_s (r+N)}{4N},\\
		\notag &\epsilon_s=\sqrt{ \lb 2s+ 2 (-1)^s m_s - 1 \rb^2-64N^2\lambda^2},\\
		\notag &\alpha_s= -\frac{1}{2} \lb s+  (-1)^s m_s \rb + \frac{1}{4} -\frac{\epsilon_s^2+32N^2\lambda^2}{4\epsilon_s}
	\end{align}
	for $s=1, 2$.
\end{thm}
\pf Setting $\eta=0$ and applying the transformation
\begin{align*}
	\Phi_{1}&=\frac{(r+N)^{\frac{m_1}{2}-\frac{5}{4}}}{e^{\frac{\epsilon_1(r+N)}{8N}}}w_1(z_1(r)),\\
	\Phi_{2}&=\frac{(r+N)^{-\frac{m_2}{2}-\frac{7}{4}}}{e^{\frac{\epsilon_2(r+N)}{8N}}}w_2(z_2(r)),
\end{align*}
where $z_s (r)$ for $s= 1, 2$ are given by \eqref{z}.
Then \eqref{r} gives
\begin{align*}
	z_1w_1''(z_1)+ &\lb m_1-\frac{1}{2}-z_1
	\rb w_1'(z_1)- \alpha_1 w_1(z_1)=0,\\
	z_2w_2''(z_2)- & \lb m_2+\frac{3}{2}+z_2
	\rb w_2'(z_2)
	-\alpha_2 w_2(z_2)=0.
\end{align*}
Using  \eqref{Kummer},  we obtain 
\begin{align*}
	w_1(r)&=\,_1F_1\lb  \alpha_1;
	m_1-\frac{1}{2}  ;z_1(r)\rb,\\
	w_2(r)&=\,_1F_1\lb  \alpha_2;
	-m_2-\frac{3}{2}  ;z_2(r)\rb,
\end{align*}
	thus we get the solutions $\Phi_{1}$ and $ \Phi_2$.  Furthermore,  \eqref{r} also gives 
\begin{align*}
	\Phi_3= &\frac{1}{\lambda} \sqrt{\frac{r +N}{r -N}} \lb\frac{(1- 2m_1) r + (2m_1 -9) N }{8N (r+N) } \Phi_1 -  \frac{\d \Phi_1}{\d r}\rb,\\
	\Phi_4= & \frac{1}{\lambda} \sqrt{\frac{r +N}{r -N}} \lb\frac{(2m_2 +3) r -(2m_2 +11)N }{8N (r+N) } \Phi_2 - \frac{\d \Phi_2}{\d r}\rb.
\end{align*}
Then by \eqref{dKummer}, we get the solutions $\Phi_3$ and $\Phi_4$.   
Therefore the theorem follows. \qed

	\mysection{Separation of the Rarita-Schwinger equation}
	\ls
	In this section, we separate the Rarita-Schwinger equation \eqref{RS} on the metric \eqref{TNT} into angular and radial equations with the ansatz 
	\begin{align}\label{Form}
		\Psi_i=  e^{\I 3(m+ \frac{1}{2})\phi}
		\begin{pmatrix}
			e^{\I \frac{3}{2} \lb m_1 +\frac{1}{2} \rb \psi} \Phi_{k1}(r) J_{k+}(\theta)\\
			e^{\I \frac{3}{2}\lb m_2 +\frac{1}{2} \rb \psi} \Phi_{k2}(r) J_{k-}(\theta)\\
			e^{\I \frac{3}{2} \lb m_1 +\frac{1}{2} \rb \psi} \Phi_{k3}(r) J_{k+}(\theta)\\
			e^{\I \frac{3}{2} \lb m_2 +\frac{1}{2} \rb \psi} \Phi_{k4}(r) J_{k-}(\theta)
		\end{pmatrix},\;i=1, 2, 
	\end{align}
	under the conditions
	\begin{align}\label{componentForm}
		\Psi_4 =e^4 \cdot   e^1 \cdot  \Psi_1,\quad    \Psi_3 =e^3 \cdot e^2 \cdot  \Psi_2,
	\end{align}
	where  $m, m_1, m_2$ are integers. 
	
	The Rarita-Schwinger equation on the metric \eqref{TNT} can be written as
	\begin{align*}
		\begin{split}
			D\Psi_{1}+\frac{f'}{f^2}e^4 \cdot \Psi_{4} + \frac{1}{2} e^1 \cdot \wt{\Psi} = \lambda \Psi_{1},\\
			D\Psi_{2} +\frac{ re^2 \cdot \Psi_{1}- N e^4 \cdot \Psi_{3} - N e^3 \cdot \Psi_{4}}{(r^2-N^2)f} +\frac{f}{2N}e^4 \cdot \Psi_{3}+ \frac{1}{2} e^2 \cdot \wt{\Psi} = \lambda \Psi_{2},\\
			D\Psi_{3} +\frac{ re^3 \cdot \Psi_{1}+ N e^4 \cdot \Psi_{2} + N e^2 \cdot \Psi_{4}}{(r^2-N^2)f} -\frac{f}{2N}e^4 \cdot \Psi_{2}+ \frac{1}{2} e^3 \cdot \wt{\Psi} = \lambda \Psi_{3},\\
			D\Psi_{4}-\frac{f'}{f^2}e^4 \cdot \Psi_{1}  + \frac{1}{2} e_4 \cdot \wt{\Psi} = \lambda \Psi_{4},
		\end{split}
	\end{align*}
	where
	\begin{align*}
		\wt{\Psi} =& - \sum_{i=1}^{4}   e_{i} \big(\Psi_{i}\big) -\lb \frac{2r}{(r^2-N^2) f} -\frac{f'}{f^2}  \rb \Psi_{1}\\
		&+\lb \frac{f'}{2f^2} e^1 \cdot e^4 -\lb \frac{N}{2(r^2-N^2) f} -\frac{f}{4N}\rb e^2 \cdot e^3\rb \cdot \Psi_4.
	\end{align*}
The separation of the above equations via \eqref{componentForm} yields the angular equations
\begin{align}\label{J}
	\begin{split}
		 \partial_{\theta} J_{i +} &=\lb  3 \lb m+\frac{1}{2} \rb  \csc \theta
		-\frac{3}{2} \lb m_1+\frac{1}{2} \rb\cot\theta \rb  J_{i +} , \\
		\partial_{\theta} J_{ i-}  &=  \lb  -3 \lb m+\frac{1}{2}\rb\csc\theta
		+\frac{3}{2} \lb m_2+\frac{1}{2} \rb\cot\theta \rb  J_{ i-} 
	\end{split}
\end{align}
for $i=1, 2$, as well as the radial equations
\begin{align}\label{Phi1}
	\begin{split}
		\partial_r \Phi_{11}=\lb \frac{3 f'}{2f} +\frac{3N}{2(r^2-N^2)} +\frac{f^2}{4N}  - \frac{3f^2}{4N}\lb m_1 + \frac{1}{2}\rb \rb \Phi_{11},\\
		\partial_r \Phi_{12}=\lb \frac{3 f'}{2f} + \frac{3N}{2(r^2-N^2)} +\frac{f^2}{4N} +  \frac{3f^2}{4N}\lb m_2 + \frac{1}{2}\rb \rb \Phi_{12},\\
		\partial_r \Phi_{13}=\lb  \frac{3 f'}{2f} -\frac{3N}{2(r^2-N^2)} -\frac{f^2}{4N} + \frac{3f^2}{4N}\lb m_1 + \frac{1}{2}\rb \rb \Phi_{13},\\
		\partial_r \Phi_{14}=\lb \frac{3 f'}{2f} -\frac{3N}{2(r^2-N^2)} -\frac{f^2}{4N} - \frac{3f^2}{4N}\lb m_2 + \frac{1}{2}\rb \rb \Phi_{14},
	\end{split}
\end{align}
and
\begin{align}\label{Phi2}
	\begin{split}
		\partial_r\Phi_{21} =\lb  \frac{f'}{2f}-\frac{2r+N}{2(r^2-N^2)}-\frac{3f^2}{4N} \lb m_1-\frac{1}{2}    \rb  \rb\Phi_{21} &-\lambda f\Phi_{23},\\
		\partial_r\Phi_{22}=\lb  \frac{f'}{2f}-\frac{2r+N}{2(r^2-N^2)}+\frac{3f^2}{4N} \lb m_2+\frac{3}{2}\rb  \rb\Phi_{22} &-\lambda f\Phi_{24},\\
		\partial_r\Phi_{23}=\lb  \frac{f'}{2f}-\frac{2r-N}{2(r^2-N^2)}+\frac{3f^2}{4N} \lb m_1-\frac{1}{2}\rb  \rb\Phi_{23} &+\lambda f\Phi_{21},\\
		\partial_r\Phi_{24}=\lb \frac{f'}{2f}-\frac{2r-N}{2(r^2-N^2)}-\frac{3f^2}{4N} \lb m_2+\frac{3}{2}\rb \rb\Phi_{24} &+\lambda f\Phi_{22}.
	\end{split}
\end{align}    
Moreover, there is a constraint
\begin{align}\label{eqPhi1}
	\lambda \Psi_{1} =0.
\end{align}
	
	\begin{thm}\label{TNTRSthetasolution}
		If $\lambda=0$, solutions of the angular equations \eqref{J} are
		\begin{align}\label{angular-solution}
			\begin{split}
				J_{1+}=J_{2+}&=\lb \sin\frac{\theta}{2} \rb^{3\lb m-\frac{m_1}{2} +\frac{1}{4}\rb}  \lb \cos\frac{\theta}{2} \rb^{
					-3\lb m+\frac{m_1}{2}+\frac{3}{4}\rb},\\
				J_{1-}=J_{2-}&=\lb \sin\frac{\theta}{2} \rb^{3\lb\frac{m_2}{2}-m-\frac{1}{4}\rb} \lb \cos\frac{\theta}{2} \rb^{3\lb m+\frac{m_2}{2}+\frac{3}{4}\rb},
			\end{split}
		\end{align} 
		which are regular if
		\begin{align}\label{RSmgeq0}
			m\geq 0,\quad 
			m_1 \leq  -2m-\frac{3}{2},\quad
			m_2 \geq  2 m+\frac{1}{2} ,
		\end{align}
		or 
		\begin{align}\label{RSm<0}
			m < 0, \quad m_1 \leq  2m+\frac{1}{2} ,\quad
			m_2 \geq  -2m-\frac{3}{2}.
		\end{align}
	\end{thm}
	\pf Solving  \eqref{J} by direct integration, we obtain \eqref{angular-solution}. The regularity follows if the exponents of $\sin{\frac{\theta}{2}} $ and $\cos \frac{\theta}{2}$ are nonnegative.  \qed
	
	\begin{rmk}
	 If  $\lambda \neq 0$, \eqref{eqPhi1}  implies  $\Psi_1=0$, and hence $J_{1\pm}$ vanish. Solutions for  $J_{2\pm}$ are still given by \eqref{angular-solution}.
	\end{rmk}
	 
	Next we solve the radial equations  \eqref{Phi1} and \eqref{Phi2} for $\lambda =0$.
	
	\begin{thm}\label{TNTRSrsolution}
		Let $g$ be the scalar flat Taub-NUT type metric \eqref{TNT} with $f$ given by \eqref{TNTf} and $C_2 > -N^2 -N C_1$.  Suppose
		\begin{align*}
			\lambda=0.
		\end{align*}
		Nonzero	solutions  of the radial equations  \eqref{Phi1} and \eqref{Phi2} on $ r\geq N$ are 
		\begin{align}\label{Phi1solution}
			\begin{split}
				\Phi_{13} &= \frac{ (r+N)^{ \frac{3}{2} } (r- r_0)^{ \frac{(6m_1+1) (r_0^2-N^2) }{8N (2r_0+C_1)} - \frac{3}{4}} e^{ \frac{(6m_1+1) r}{8N}} }
				{  (r+r_0+C_1)^{-\frac{(6m_1+1)\lb N^2-(r_0+C_1)^2\rb}{8N(2r_0+C_1)}+\frac{3}{4}}},\\
				\Phi_{14}&=\frac{ (r+N)^{\frac{3}{2}}(r-r_0)^{-\frac{(6m_2+5)(r_0^2-N^2)}{8N(2r_0+C_1)}-\frac{3}{4}}e^{-\frac{(6m_2+5)r}{8N}}}{ (r+r_0+C_1)^{\frac{(6m_2+5)\lb N^2-(r_0+C_1)^2\rb}{8N(2r_0+C_1)}+\frac{3}{4}}},
			\end{split}
		\end{align}
		and
		\begin{align}\label{Phi20solution}
			\begin{split}
				\Phi_{23}&=\frac{(r- r_0)^{ \frac{(6m_1-3)(r_0^2 -N^2)}{8N(2r_0 +C_1)} -\frac{1}{4} }  e^{\frac{(6m_1-3) r}{8N}}}
				{ (r+N)^{\frac{1}{2}}   (r+r_0+C_1)^{-\frac{(6m_1-3)
				\lb N^2-(r_0+C_1)^2\rb }{8N (2r_0+C_1)} +\frac{1}{4}} },\\
				\Phi_{24}&=\frac{ (r-r_0)^{-\frac{(6m_2+9)(r_0^2-N^2)}{8N (2r_0+C_1)}
						-\frac{1}{4}}  e^{-\frac{(6m_2+9)r}{8N}} }
				{ (r+N)^{\frac{1}{2}} (r+r_0+C_1)^{\frac{(6m_2+9)
				\lb N^2-(r_0+C_1)^2 \rb}{8N (2r_0+C_1)} +\frac{1}{4}} }.
			\end{split}
		\end{align}
	Solutions of the angular equations \eqref{J} are given by  \eqref{angular-solution}.  Moreover, under the conditions \eqref{RSmgeq0} or \eqref{RSm<0}, these solutions are $L^2$ integrable. 
	\end{thm}
	\pf
    Solving \eqref{Phi1} and \eqref{Phi2} with $\lambda=0$, we obtain the solutions \eqref{Phi1solution} and \eqref{Phi20solution}.
    Under the conditions \eqref{RSmgeq0} or \eqref{RSm<0} that ensure $m_1\leq -1$ and   $m_2 \geq 0$ in \eqref{Form}, the expression
	 	\begin{align*}
	 	(r^2 - N^2) \sum_{i=1}^{2} \lb  \lv \Phi_{i3}\rv^2 +   \lv \Phi_{i4}\rv^2 \rb 
	 \end{align*}
	  decays to zero as  $r \rightarrow \infty$.  Thus, there exists a constant $C'$ such that
	\begin{align*}
		\int_{\mathcal{D}} \int_{ N}^{\infty}\left| \Psi_k \otimes e^k \right|^2 \d\mu 
		<  C'  \int_{N}^{\infty}  (r^2-N^2) \sum_{i=1}^{2} \lb  \lv \Phi_{i3}\rv^2 +   \lv \Phi_{i4}\rv^2 \rb  \d r < \infty.
	\end{align*}
	This completes the proof.    \qed

	\begin{rmk}
	    We consider Rarita-Schwinger fields satisfying \eqref{rsfeild} and \eqref{divergence} on the metric \eqref{TNT}, where $f$ is given by \eqref{TNTf} with $C_2 > -N^2 -N C_1$. 
	    Solutions of \eqref{rsfeild} taking the form \eqref{componentForm} are given by Theorems \ref{TNTRSthetasolution} and \ref{TNTRSrsolution}, and the additional constraint \eqref{divergence} requires that  $J_{1\pm}$ in \eqref{angular-solution} and $\Phi_{1j}$ $(j=3,4)$ in \eqref{Phi1solution} vanish.
	\end{rmk}
	
	For  $\lambda \neq 0$, \eqref{eqPhi1}  implies  $\Psi_1=0$, and we now solve the radial equations \eqref{Phi2} of $\Psi_2$.

	\begin{thm}
		Let $g$ be the scalar flat Taub-NUT type metric \eqref{TNT} with $f$ given by \eqref{TN}. Suppose
		\begin{align*}
			\lambda \neq 0.
		\end{align*}
		If $ m_1 \leq -1$ and  $m_2 \geq 0 $ in \eqref{Form},  solutions of the radial equations \eqref{Phi2} on $r\geq N$ are
		\begin{align*}
			\begin{split}
				\Phi_{21}=&\frac{(r-N)^{-\frac{3m_1}{2}-\frac{1}{4}}}{\sqrt{e^{z_1}}} 
				\,_1F_1\lb  \alpha_1;	\frac{3}{2}-3m_1 ; z_1 \rb,\\
				\Phi_{22}=&\frac{(r-N)^{\frac{3m_2}{2} +\frac{5}{4}}}{\sqrt{e^{z_2}}} \,_1F_1\lb  \alpha_2;  3m_2+\frac{9}{2}  ; z_2 \rb,\\
				\Phi_{23}=&\frac{(r-N)^{-\frac{3m_1}{2} +\frac{1}{4}}}{8N \lambda \sqrt{(r+N) e^{z_1}} }\lb (3- 6m_1- \epsilon_1)\,_1F_1\lb  \alpha_1;
				\frac{3}{2} -3m_1 ; z_1 \rb\right.\\&\left. +\frac{\epsilon_1^2 -\epsilon_1 (3-6m_1) + 32N^2 \lambda^2}{6m_1-3} 	
				\,_1F_1\lb  \alpha_1+1;
				\frac{5}{2}-3 m_1 ; z_1 \rb\rb,\\
				\Phi_{24}=&\frac{(r-N)^{\frac{3m_2}{2} +\frac{7}{4}}}{8N\lambda \sqrt{(r+N)e^{z_2}}}\lb
				(6m_2 +9-\epsilon_2)\,_1F_1\lb  \alpha_2;	3m_2+\frac{9}{2}  ;  z_2 \rb\right.\\&\left.-\frac{\epsilon_2^2 -\epsilon_2 (6m_2+9) +32N^2 \lambda^2}{6m_2+9} 	
				\,_1F_1\lb  \alpha_2+1; 3m_2+\frac{11}{2}  ; z_2 \rb\rb,
			\end{split}
		\end{align*}
		where
		\begin{align*}
			&z_s (r) =	-\frac{\epsilon_s (r -N)}{4N},\\
			&\epsilon_s =\sqrt{(3^s + (-1)^s 6m_s)^2-64N^2\lambda^2},\\
			&\alpha_s = \frac{1}{4}\lb 3^s + (-1)^s 6m_s \rb - \frac{\epsilon_s^2 + 32N^2 \lambda^2}{4\epsilon_s}
		\end{align*}
		for $s=1, 2$.
	\end{thm}
	\pf  The theorem can be proved by using the same argument as the proof of Theorem \ref{DiracTN-Nthm4.5}. \qed
	
	\bigskip
	
	{\footnotesize {\it Acknowledgement. 
		This work was supported by National Natural Science Foundation of China (Grant No.12301072).  The authors are grateful to Professor Zhang Xiao for his invaluable advice on this paper.

	\end{document}